# Modification of electronic structure and morphology of Ni$_2$MnGa(100) by Cr adlayers

Abhishek Rai+, Jayita Nayak+

UGC-DAE Consortium for Scientific Research, Khandwa Road, Indore, 452001, India

+Both the authors have contributed equally to the work

## Abstract

The growth of Cr adlayers on the Mn-Ga terminated surface of Ni$_2$MnGa(100) has been studied in this work. We show formation of islands of epitaxial Cr layer up to 4 atomic layer heights using scanning tunneling microscopy, while core-level x-ray photoelectron spectroscopy (XPS) indicates absence of any intermixing of the adlayer and the substrate. The density of states (DOS) of the thin Cr layer (2.7 ML) measured by scanning tunneling spectroscopy exhibits nice agreement with density functional theory (DFT) that establishes that it is ferromagnetic. In contrast, a thick layer of Cr that also grows epitaxially on Ni$_2$MnGa(100) is shown to be antiferromagnetic by comparing its XPS valence band with the DOS for bulk Cr calculated by DFT.

# INTRODUCTION

Ultra thin adlayers of antiferromagnetic materials on ferromagnetic substrates have been studied extensively in recent years because localized surface and interface magnetic states offer the possibility of inducing interesting phenomena and hold out the promise of device applications specifically for giant magnetoresistance (GMR), spin valves and spin transfer torque memory devices. Enhancement of perpendicular magnetic anisotropy and its electric field-induced change through interface engineering is possible in Cr/Fe/MgO layer, which is very useful for magnetic memory devices [1]. Cr on a ferromagnetic substrate exhibits ferromagnetism in the topmost surface layer [2], although it is antiferromagnetic in the bulk. Moreover, Cr exhibits incommensurate spin density wave along [100] direction in the bulk with a wavelength of approximately 21 lattice spacings. Recently, Rzhevsky *et al.* observed that ultrashort optical pulses can induce long-living magnetization oscillations in Cr/Fe (100) thin films due to a thermal mechanism of the excitation as studied by optical pump-probe approach [3]. In this context, the non-collinear magnetic moment distribution at the Cr/Fe(110) interface of an Fe(110) thin film covered by a Cr monolayer was predicted by periodic Anderson model calculations [4] which was recently confirmed by spin resolved APRES indicating the dependence on film thickness to a rapidly decreasing and oscillating photoelectron spin polarization [5]. From theoretical calculations for Cr, Hirashita *et al.* showed that the magnetic moment is largest for the first layer and decreases substantially for the second and third layers[6]. Recently, there is special interest in Cr/Fe systems, where the top layer, due to its antiferromagnetism in the bulk, may play an active role in mediating the exchange coupling. The antiparallel coupling across the interface and a strong enhancement of the Cr surface magnetic moment, with respect to the bulk, were predicted for an ordered Cr monolayer on Fe(100) [7, 8]. Thin films of Cr grown on Fe(100) reveal thickness-dependent short-period oscillations of the photoemission intensity close to the Fermi energy and the oscillations are assigned to quantum-well states caused by the nesting between the Fermi-surface sheets around the Γ and the X points in the Brillouin zone of antiferromagnetic Cr[9].

$Ni_2MnGa$ belongs to a class of active material namely ferromagnetic shape memory alloy. In ferromagnetic shape memory alloys, very large strain up to 10% has been reported by the application of magnetic field[10]; rendering it to be an important material for actuator and sensor applications[11, 12]. At room temperature, Ni2MnGa has a cubic L21 structure. It undergoes a symmetry lowering transformation to the martensite phase around 200 K to a modulated phase with orthorhombic unit cell[13, 14]. Ni2MnGa is ferromagnetic at room temperature with a large magnetic moment (4 µB) primarily localized on Mn[13]. Thus, Cr on Ni2MnGa is expected to exhibit exotic physical properties. Epitaxial growth is expected since the lattice constant of Ni2MnGa in the austenite phase (0.582 nm) is double of the lattice constant of bcc Cr (0.288 nm). Moreover, existence of charge density wave in Ni2MnGa[15] and spin density wave in Cr has been reported. This indicates the importance of studying the electronic structure and morphology of Cr/Ni2MnGa.

## METHODS

STM measurement was performed in a VT-STM system from Omicron GmbH, Germany at a base pressure of about $6\times 10^{-11}$ mbar. All the measurements were performed using tungsten tips that were cleaned by voltage pulse method. The images were recorded in the constant current mode with the specimen at ground potential. XPS was performed using Al $K_\alpha$ x-ray source (hγ= 1486.6 eV) and Phoibos 100 electron energy analyzer from Specs GmbH, Germany. $Ni_2MnGa$(100) crystal was grown by Bridgman method, subsequently cut along [100] axis in the austenite phase and then mechanically and chemically polished[16]. The stoichiometric Ni2MnGa(100) surface was prepared by sputtering of Ar+ ion with 0.5 keV energy followed by annealing at 770 K for about an hour[17]. Before Cr deposition, the surface cleanliness and order was monitored by XPS and LEED, respectively. Cr deposition has been executed by using a EFM cell at base pressure of $2\times 10^{-10}$ mbar. The ground state electronic structure of bulk Cr metal has been calculated in the space group Im3m corresponding to the body centered cubic structure with the lattice parameter a= 0.288 nm, as obtained from the present STM measurements. These calculations havebeen performed by using scalar relativistic full potential spin polarized Korringa- Kohn- Rostoker method based on the Greens function formalism as implemented in the SPRKKR code[18]. The exchange and correlation effects were incorporated within the generalized gradient approximation framework[19]. The angular momentum expansion up to lmax=4 has been used for the multipole expansion of the Greens function. The Brillouin zone integrations were performed on a 28×28×28 mesh of k-points in the irreducible wedge of the Brillouin zone. The energy convergence criterion has been set to $10^{-5}$ Ry.

Sharp four-fold LEED pattern of the (100) surface (Fig. 1(a)) represents that the surface is well ordered and free of contamination. The STM topographic image of Ni$_2$MnGa(100) surface in the austenite phase (Fig. 1(b)) shows atomically flat terraces separated by an atomic step. In Fig. 1(c), a height profile along the black line marked by a-b in Fig. 1(b) shows that the step height is 0.29±0.02 nm. Interestingly, this is half of the bulk lattice parameter (a= 0.582 nm) of the L21 structure. The L21 structure is best described as four inter penetrating f.c.c. sublattices with Mn sublattice at (0, 0, 0), Ga at (0.5, 0.5, 0.5), and Ni at (0.25, 0.25, 0.25) and (0.75, 0.75, 0.75); the separation between two Mn-Ga layers or two Ni layers is 0.291 nm. The step height of 0.29 nm obtained from STM shows that the surface is bulk terminated and both the upper and the lower terraces have same type of termination, which is either Mn-Ga or Ni type. Fig. 1(d) presents atomically resolved STM image of Ni$_2$MnGa(100) surface in the austenite phase and the corresponding FFT image (inset Fig. 1(d)). This provides further insight into the surface termination. A regular four fold arrangement of the atoms of the surface is clearly observed. A few dark holes indicate defects, such as vacancies.

The lattice parameter obtained from the atomically resolved image is 0.42±0.02 nm, which is similar to the lattice parameter obtained by Leicht et. al [20]. This value is close to the bulk lattice parameter corresponding to the primitive unit cell of the L21 lattice given by a/ɣ2= 0.412 nm[21]. Thus, the lattice constant of the surface and the bulk are quite similar indicating bulk termination, as reported from our earlier studies[22]. A significance of the present value of the surface lattice parameter (0.42 nm) is that the surface is Mn-Ga terminated. This is because, if the surface would have been Ni terminated, the expected lattice parameter would have been 0.29 nm. In fact, Mn-Ga termination of the Ni$_2$MnGa(100) surface was previously reported by Dhaka et al. from angle dependent x-ray photoemission studies[17] as well as for Ni-Mn-Ga(100) epitaxial thin films grown on MgO(100)[23].

The Cr 2p core-level spectra as a function of coverage is shown in Fig. 2(a). The Cr $2p_{3/2}$ peak at 574.05 eV gradually increases in intensity with coverage, while its full width at half maximum (FWHM) is reduced for higher coverages. It may be noted that the broad features at 587.7 eV and 570 eV that are observed at low Cr coverages are related to Ga L$_3$M$_{23}$M$_{23}$ Auger main peak and related satellite features from the substrate. Ni $2p_{3/2}$ and $2p_{1/2}$ spectra of Ni2MnGa substrate appear at 852.7 eV and 870 eV binding energies, respectively (Fig. 2(b)). Fig. 2(c) shows a gradual increase (decrease) in the adlayer (substrate) related intensity of the Cr 2p (Ni 2p) with coverage. The Ni 2*p* spectrum of Ni$_2$MnGa exhibits a satellite at 6.7 (5.7) eV higher binding energy from Ni 2p3/2 (Ni 2p1/2) peak. These satellites have been explained in literature to have similar origin as the well known 6 eV satellites in Ni metal[25]. In Ni 2p core-level spectra, the satellite at 6 eV from the main peak arises from the interaction of d states with the s states[25]. We find that deposition of Cr does not affect these satellite peaks both in terms of their position and relative intensity w.r.t. the Ni 2p main peak. Moreover, the main peak positions of both the adlayer and substrate remain unchanged with coverage, and no additional features emerge. These observations indicate absence of any intermixing or alloying between Cr and the substrate.

The STM topographic image of 2 ML Cr/Ni$_2$MnGa(100) provides the evidence of Cr islands (Fig. 3(a)), although the corresponding LEED pattern exhibits sharp spots with four fold symmetry (Fig. 3(c)). In order to find out the distribution of the island heights, we show a histogram in Fig. 3(b). From the histogram, the interesting observation is that the islands mostly acquire two different step heights: 0.3 nm, 0.43 nm (black arrows), and smaller number of islands have heights of 0.16 and 0.57 nm (red arrows). Since the lattice constant of bulk Cr that has b.c.c. structure is 0.29 nm, the island heights of 0.16, 0.3, 0.43 and 0.57 nm nicely correspond to first to fourth Cr(100) layer. Thus, the islands grow epitaxially, and the STM result supports the conclusion of epitaxial growth from LEED. Somewhat similar situation has been observed for Cr/Au(111) where even before the bare Au surface is fully covered, the second layer growth starts through formation of two and three atomic step height Cr islands[26, 27]. This indicates that the growth is of Stranski- Krastanov type.

Scanning tunneling spectroscopy (STS) measurement was also performed for Ni$_2$MnGa(100) that shows two peaks at -0.6 eV in the unoccupied region i.e. above E$_F$ and at 0.4 eV in the occupied region i.e. below EF (green up triangles in Fig. 4(a)). In order to ascertain their origin, we compare them with the density of states (DOS) of Ni2MnGa, calculated using full potential linearized augmented plane wave method in our earlier work[24]. The peak at -0.6 eV in STS arises due to hybridized Ni 3d-Mn 3d states. However, there is a shift of 0.3 eV between experiment and theory, which is possibly related to the surface effects, since the calculation was performed for the bulk. The peak at 0.4 eV has a close resemblance to feature A′ in the DOS of Ni2MnGa (Fig. 2 of Ref. 24) that arises primarily from Ni 3d states.

The STS spectrum of Ni2MnGa(100) discussed above is compared with that of 2.7 ML

Cr/Ni$_2$MnGa(100) (black filled circles) in Fig. 4(a). The Cr STS spectrum exhibits four prominent peaks at -1.07 eV, -0.1 eV, 0.65 eV and 1.2 eV. In order to determine the origin of these peaks, we have calculated the DOS of the bulk Cr using SPR-KKR code that converges to an antiferromagnetic solution. However, in the calculated bulk Cr DOS, we find only two features - one at 1.3 eV below the Fermi level and another at -1.4 eV i.e. above Fermi level. Thus, the features observed in the STS spectra cannot be explained from the bulk Cr DOS. Since the Cr monolayer has been reported to be ferromagnetic in literature, we have compared our STS spectrum with the spin polarized DOS calculation for 1 ML Cr(100) on Fe(100) substrate from Fig. 7(a) of Ref. 7. This theoretical calculation shows that the Cr layer is ferromagnetic, the valence band being dominated by the minority states, while the conduction band is dominated by the majority states resulting in large net moment[7]. It is interesting to note that all the peaks shown by colored arrows in the STS spectrum are reproduced nicely by the DOS for ferromagnetic 1 ML Cr(100) (corresponding to the arrows of same color)(Fig. 4(b)). This nice agreement of the STS spectrum with ferromagnetic Cr(100) monolayer clearly indicates that the top Cr layer on the ferromagnetic Ni2MnGa substrate is also ferromagnetic.

Turning to the thick layer (_80 ML) of Cr on Ni$_2$MnGa deposited at room temperature, we note that the presence of sharp LEED spots demonstrate that the growth is epitaxial (Fig. 5)(a). The LEED pattern shows that although the Cr layer retains its fourfold symmetry (shown by the pink dotted line), the spot positions exhibit R45∘ rotation compared to Ni$_2$MnGa(100) (Fig. 1(a). Since the lattice constant of Ni$_2$MnGa(100) (0.582 nm) is about two times of the lattice constant of Cr(100) (0.288 nm), it is expected that the Cr layer is oriented along (100) direction. The presence of a fourfold symmetry rules out possibility of (110) [(111)] orientation since in that case the LEED pattern would be rectangular [hexagonal].

The spectral shape of the XPS valence band of the Cr thick layer is differs from the substrate Ni$_2$MnGa[28] with appearance of new features. The valence band of the thick Cr layer exhibits features such as A at 1.3 eV, B at 2.4 eV and C at 3.2 eV (Fig. 5(b)). We have compared the XPS VB with the theoretical DOS calculated using the SPRKKR code. Unlike for STS of the thin Cr layer, here the agreement with the bulk antiferromagnetic calculation is fairly good. The features A and B arise mainly due to Cr *d* band, with a small contribution from 4*p* state. On the other hand, feature C originates from the combined contribution from 3*d*, 4*p*, and 4*s* states. This shows that the thick Cr layer is antiferromagnetic in contrast to the thin layer that is ferromagnetic.

CONCLUSIONS

Epitaxial Cr layers have been grown for the first time on the ferromagnetic shape memory alloy Ni$_2$MnGa(100). STM measurements reveal the tendency of formation of Cr islands up to 4 atomic layer heights, which coalesce at higher coverages. The density of states of the Cr thin layer that is dominated by Cr 3d states around the Fermi level has been studied by STS and excellent agreement with theory is obtained indicating that it is ferromagnetic. In, a thick Cr layer that also grows epitaxially on Ni$_2$MnGa(100) as Cr(100) shows good agreement with bulk DOS of antiferromagnetic Cr, indicating that the thick layer is not ferromagnetic.

ACKNOWLEDGEMENTS

We thank D. Schlagel and T. Lograsso for providing the Ni2MnGa single crystal substrate.

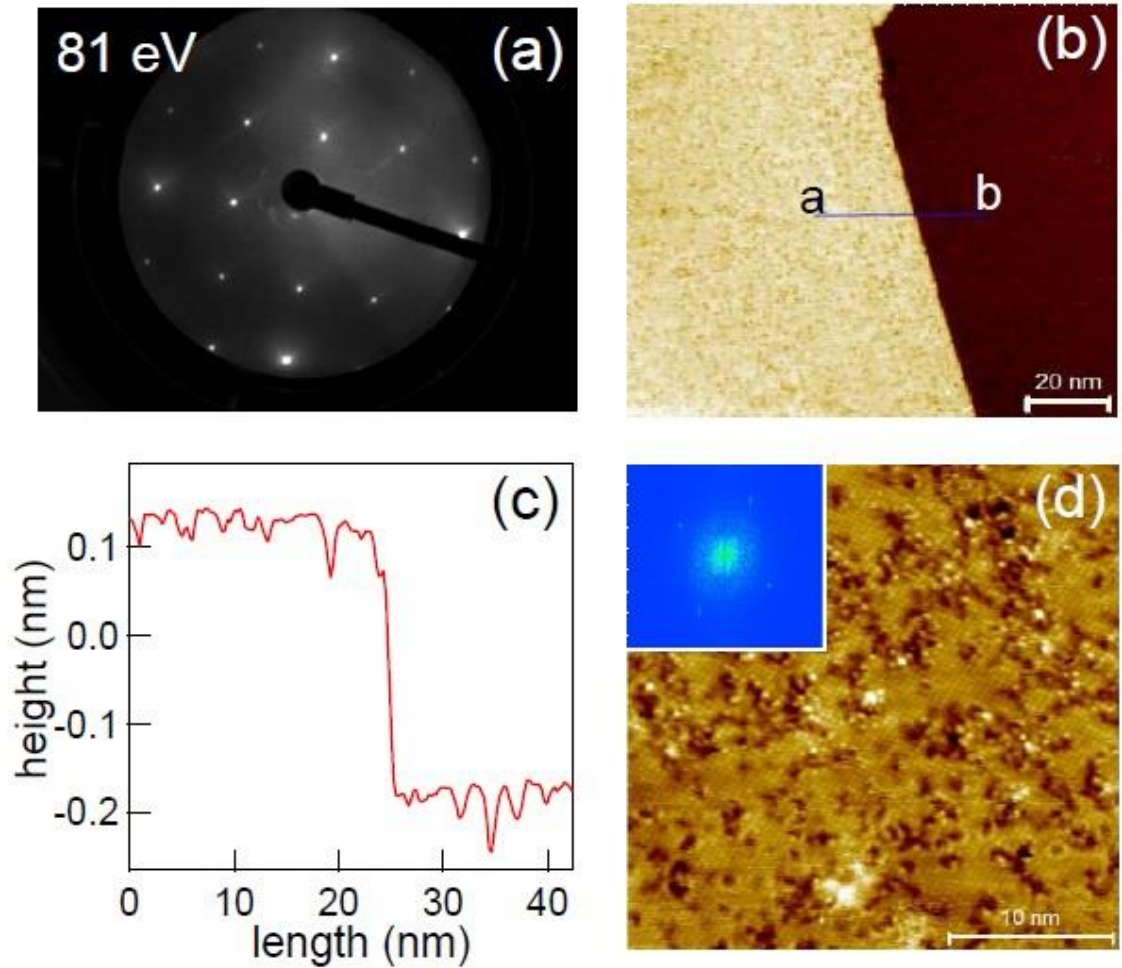

FIG. 1: (a) LEED pattern of Ni$_2$MnGa(100) surface at room temperature. The primary electron beam energy is indicated in the top left corner. (b) STM topographic image of Ni$_2$MnGa(100) surface (IT= 0.4 nA, UT= 1.6 V) at room temperature after annealing at 770 K. The image shows large uniform terraces and an atomic step. (c) The average height profile along a − b showing the step height of 0.29 nm, (d) Atomically resolved STM image (IT= 0.4 nA, UT= 0.3 V) of Ni$_2$MnGa(100) surface at room temperature and inset shows corresponding fast Fourier transform (FFT).

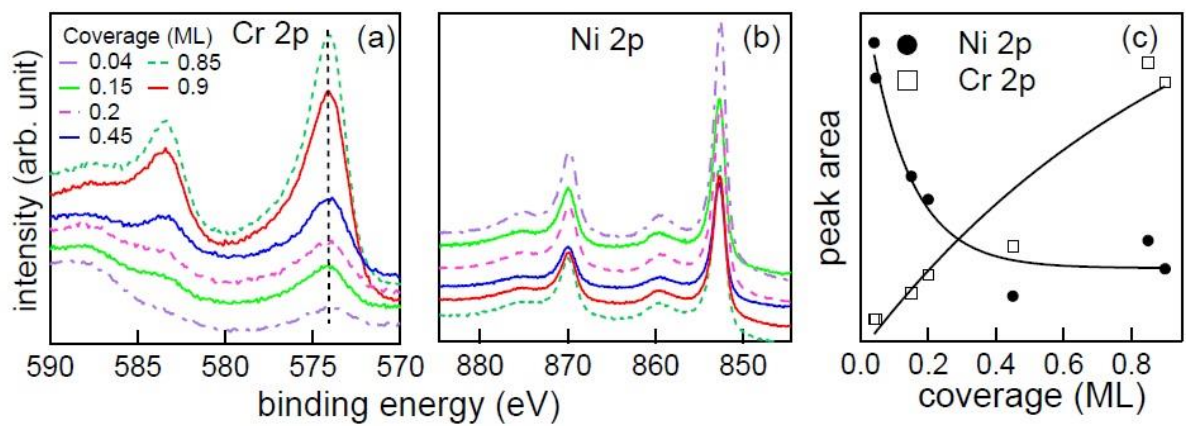

FIG. 2: (a) Cr 2p and (b) Ni 2p XPS core level spectra for Cr/Ni$_2$MnGa(100) as a function of coverage at room temperature. (c) Variation of Cr 2p and Ni 2p core-level intensities as a function of coverage, the solid lines serve as guide to the eye.

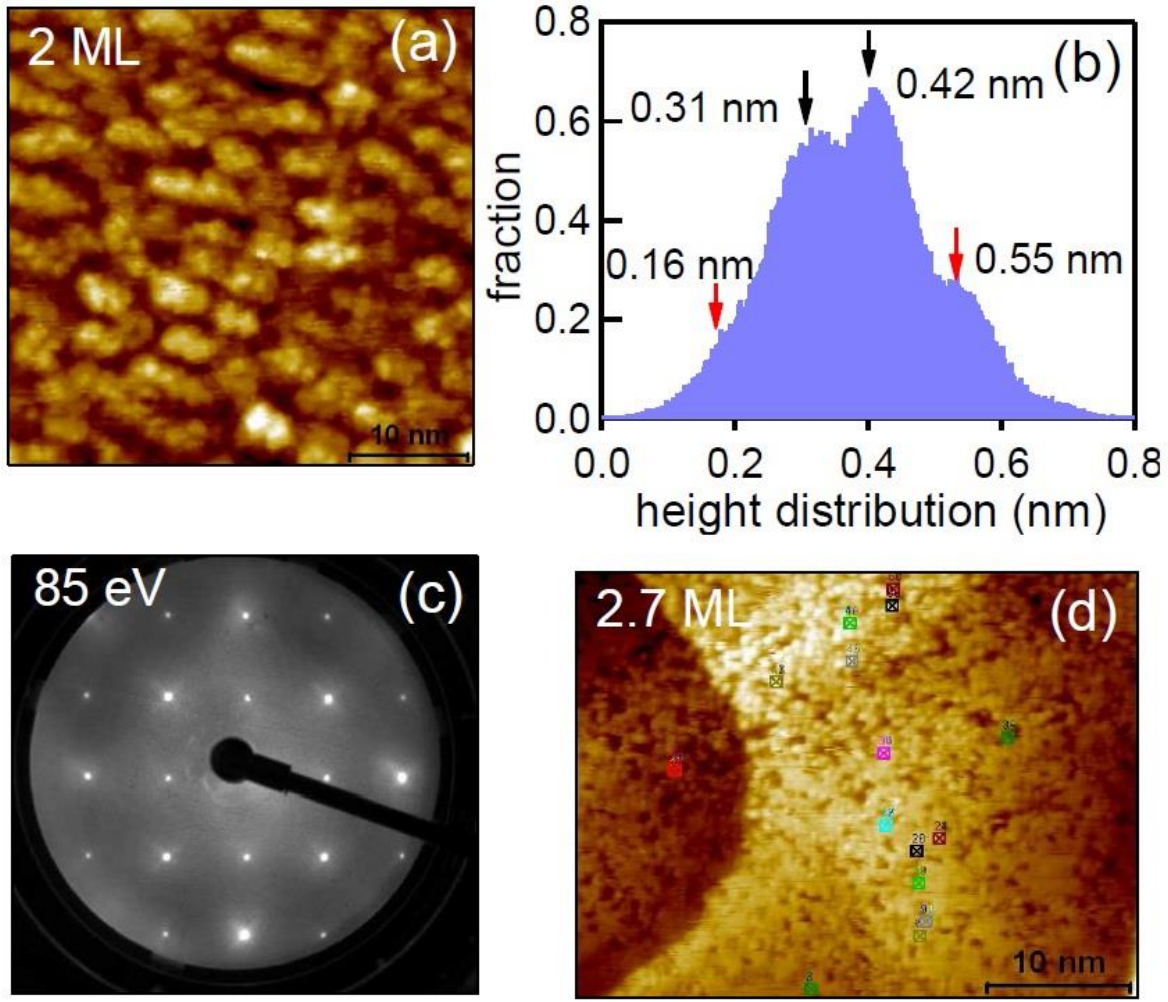

FIG. 3: (a) STM topographic image ($I_T$= 1 nA, $U_T$= 1.7 V) of 2 ML Cr/Ni$_2$MnGa(100) at room temperature and (b) histogram showing the height distribution of Cr clusters in (a), the arrows indicate the peak positions, (c) LEED pattern of 2 ML Cr deposited surface acquired with $E_p$=85 eV. (d) STM topographic image of about 2.7 ML Cr on Ni$_2$MnGa(100), the marked regions indicate the points where STS has been performed.

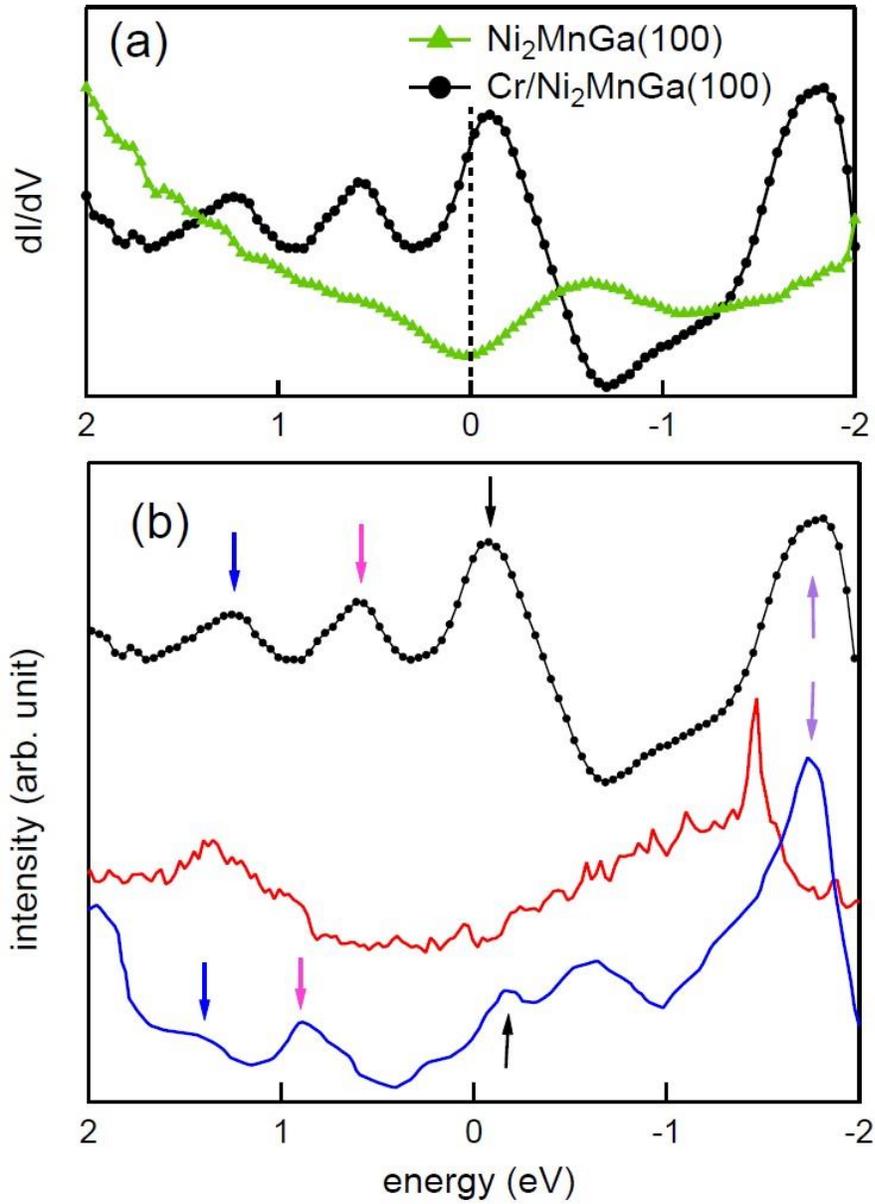

FIG. 4: (a) Comparison of STS spectra of Ni$_2$MnGa(100) and 2.7 ML Cr/Ni$_2$MnGa(100). (b) The STS spectrum of Cr/Ni$_2$MnGa(100) has been compared with theoretically calculated density of states (DOS) of bulk Cr (red line) and that of 1 ML Cr(100) on Fe(100) from Ref. 7 (blue line). The arrows of the different colors are used to identify the different features.

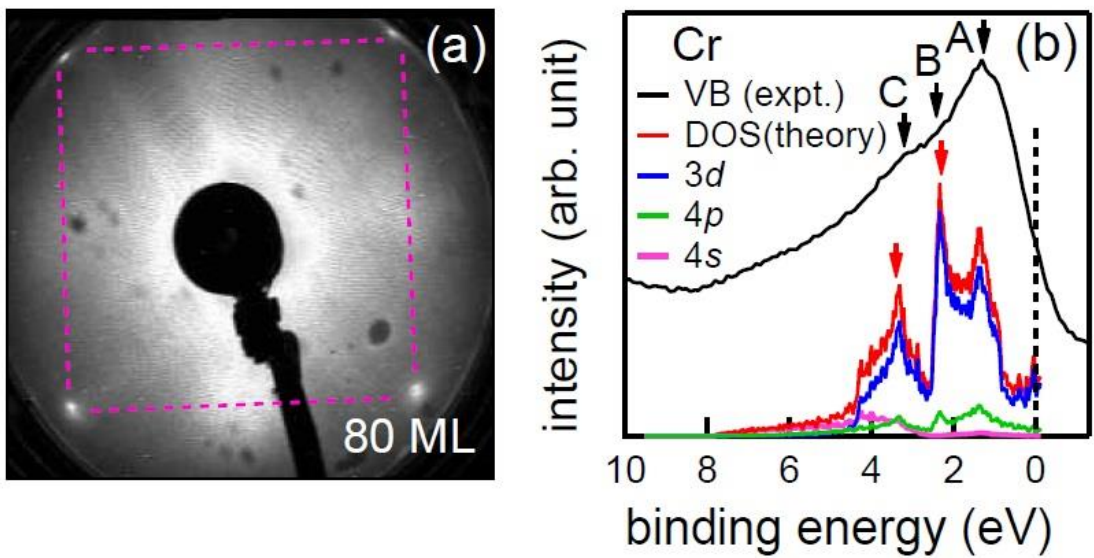

FIG. 5: (a) LEED patterns of a thick layer Cr on Ni$_2$MnGa(100) at 95 eV beam energy. (b) Comparison of the experimental valence band of Cr thick layer recorded using x-ray photoelectron spectroscopy with total and Cr 3d, 4p and 4s partial DOS calculated using density functional theory.